\documentclass[letterpaper,10pt]{article} 
\usepackage{osameet3} %% use version 3 for proper copyright statement
\usepackage{bm}
\usepackage{amsmath,amssymb, physics}
\usepackage[colorlinks=true,bookmarks=false,citecolor=blue,urlcolor=blue]{hyperref} %pdflatex
\usepackage{ulem}
\usepackage{units}

\newcommand{\ml}{l}
\newcommand\authormark[1]{\textsuperscript#1}

\begin{document}

\title{Spin-orbit interaction in nanofiber-based Brillouin scattering}
%\title{Spin-orbit coupling-induced circular polarization inversion in nanofiber-based Brillouin backscattering}

\author{Maxime Zerbib \authormark{1}, Maxime Romanet \authormark{1}, Thibaut Sylvestre \authormark{1}, Christian Wolff \authormark{2}, Birgit Stiller \authormark{3}\authormark{,}\authormark{4}, Jean-Charles Beugnot \authormark{1} and Kien Phan~Huy \authormark{1}\authormark{,}\authormark{5}}

\address{\authormark{1} FEMTO-ST Institute, UMR 6174 CNRS-Université de Franche-Comté, 25030 Besançon, France\\
\authormark{2} Center for Nano Optics, University of Southern Denmark, Campusvej 55, DK-5230 Odense M, Denmark\\
\authormark{3} Max Planck Institute for the Science of Light, Staudtstr. 2, 91058 Erlangen, Germany\\
\authormark{4} Department of Physics, University of Erlangen-Nuremberg, Staudtstr. 7, 91058 Erlangen, Germany\\
\authormark{5} SUPMICROTECH-ENSMM, 16 rue de l'épitaphe, 25000 Besançon, France \\
}

\email{maxime.zerbib@femto-st.fr}
%%Uncomment the following line to override the copyright year from the default current year.
\copyrightyear{2023}

\begin{abstract}
Angular momentum is an important physical property that plays a key role in light-matter interactions such as spin-orbit interaction. Here, we investigate theoretically and experimentally the spin-orbit interaction between a circularly polarized optical (spin) and a transverse vortex acoustic wave (orbital) using Brillouin backscattering in a silica optical nanofiber. We specifically explore the state of polarization of Brillouin backscattering induced by the TR21 torso-radial vortex acoustic mode that carries an orbital angular momentum. Using a full-vectorial theoretical model, we predict and observe two operating regimes for which the backscattered Brillouin signal is either depolarized or circularly polarized depending on the input pump polarization. We demonstrate that when the pump is circularly polarized and thus carries a spin angular momentum, the backscattered signal undergoes a handedness reversal of circular polarization due to optoacoustic spin-orbit interaction and the conservation of overall angular momentum.
\end{abstract}

\section{Introduction}
Spin-orbit interaction (SOI), whereby spin and orbital features of a particle or a wave affect each other, is a striking and ubiquitous phenomenon occurring in many fields of physics, such as solid-state and quantum physics, optics and acoustics. In optics, it has long been known that light carries both orbital angular momentum (OAM) and spin angular momentum (SAM), related to wavefront rotation and circular polarization, respectively, and that they can become coupled by SOI while conserving the overall angular momentum \cite{ZayatsNP2015}. As early as 1909, Poynting suggested that circularly polarized light possesses a SAM known as the spin of light \cite{poynting_wave_1909}. The conservation of angular momentum in an optomechanical interaction was then experimentally evidenced in 1936 by Beth \cite{beth_mechanical_1936}, who showed that the conversion of left-handed circular polarization to right-handed circular polarization through a crystalline waveplate induces an opposite torque on the waveplate. This pioneering experiment still resonates in the scientific community today, as shown by numerous studies over the past decades in optical tweezers and other applications based on SOI \cite{padgett_tweezers_2011,hakobyan_left-handed_2014}. Although this effect attracts interest beyond optics \cite{franke-arnold_optical_2017}, it is at the heart of many studies in photonics \cite{bliokh_spinorbit_2015} with major applications in the manipulation of small objects and light itself \cite{wang_magnetic_2018}. The coupling between light SAM and OAM can give rise to intriguing phenomena such as photonic spin-Hall effect \cite{PhysRevLett.93.083901,hosten_observation_2008} and spin-dependent vortex generation\cite{PhysRevLett.96.163905}. In acoustics, SOI has been recently reported using spin and angular momentum of transverse acoustic beams \cite{wang_spin-orbit_2021,PhysRevB.99.174310}. 

In this paper, we report an optoacoustic spin-orbit interaction using Brillouin backscattering in a silica optical nanofiber \cite{tong_subwavelength-diameter_2003,beugnot_brillouin_2014,godet_brillouin_2017}, whereby a circularly polarized optical mode carrying a SAM coherently interacts with a torso-radial (TR21) vortex acoustic mode that has non-zero OAM. We show in particular that the backscattered Brillouin signal by the TR21 acoustic mode has a opposite circular polarization due to the spin-orbit coupling and the conservation of total angular momentum. Our observations complete other recent studies about angular momentum conservation in both chiral photonic crystal fibers\cite{zeng_stimulated_2022,ZengVortexIsol,VortexZengArxiv2022} and in standard single-mode fibers using Forward Brillouin scattering \cite{diamandi_interpolarization_2022,bashan_forward_2021}. They are relevant not only for further fundamental investigations but also for potential applications, such as new distributed sensing schemes based on SOIs or Brillouin-based optical memories \cite{merklein_chip-integrated_2017,Stiller2020}. For the latter, it can open the opportunity to store not only the coherent information of light but also different polarization states, therefore enhancing the storage capacity. 

The paper is organized as follows. In the first section, we describe the theoretical background of Brillouin light backscattering by the TR21 acoustic vortex wave in an optical nanofiber. We then show that, when the incident light is linearly polarized, we observe a polarization scrambling of the scattered light similar to previous forward Brillouin scattering experiments \cite{shelby1985guided}. On the other hand, when the pump light is circularly polarized and therefore has a SAM, the SOI allows the backscattered light to be fully polarized with handedness reversal. In the second and third sections, we present experimental polarization-sensitive measurements of Brillouin backscattering in a 730-nm diameter silica nanofiber, and we confirm our theoretical predictions.

\section{Principle and Theory}

\begin{figure}[h]
    \centering
    \includegraphics[width=8 cm]{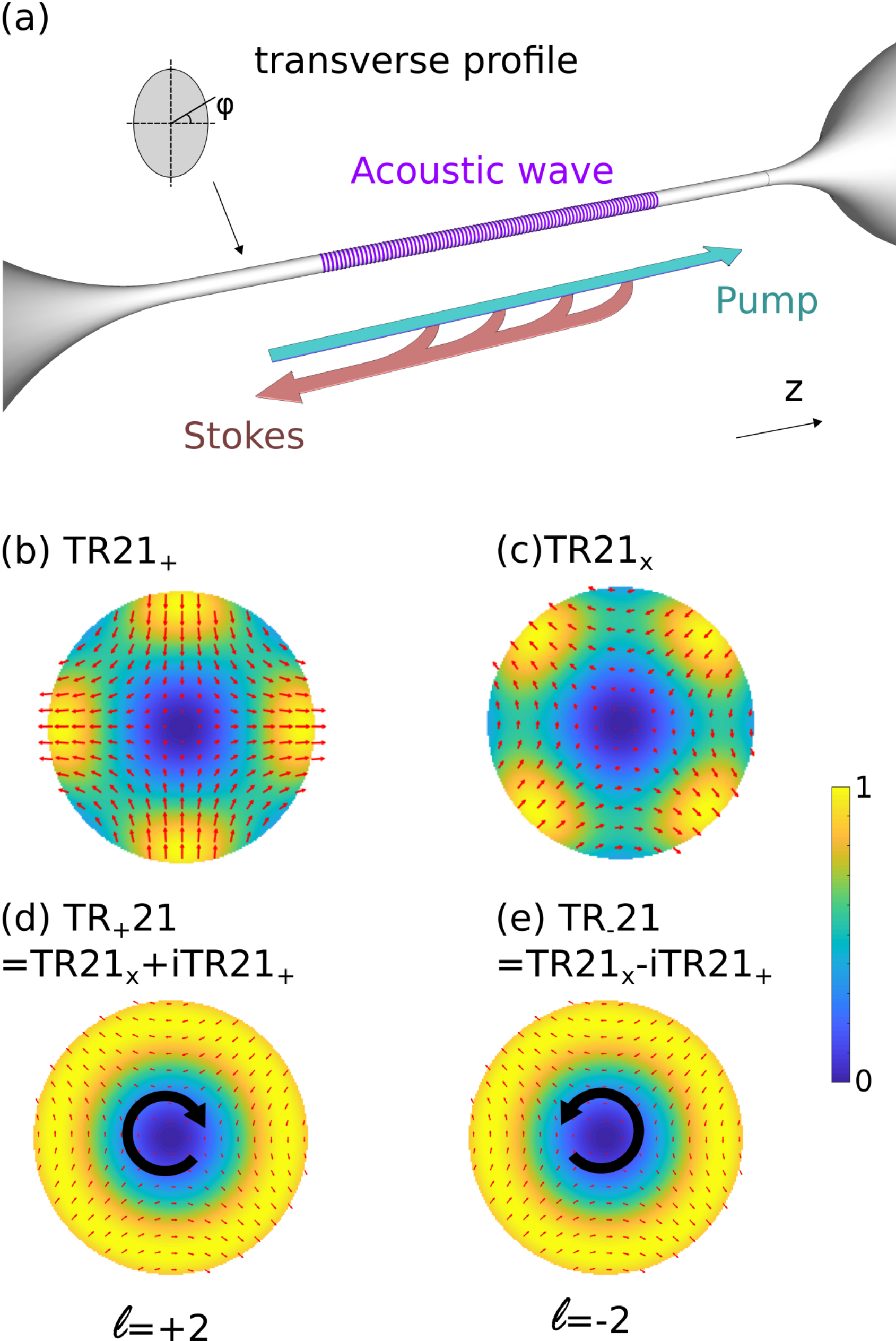}
    \caption{(a) Illustration of Brillouin backscattering in a tapered optical nanofiber. (b) and (c) are the TR21$_+$ and TR21$_\times$ degenerated stationary torso-radial modes assuming a slight elliptical core. The red arrows show the transverse displacement field and the false colors show the normalized transverse kinetic energy. (d) and (e) are the vortex acoustic modes TR+21 and TR-21 resulting from the linear combination of the two previous stationary modes.}
    \label{fig:nanofil}
\end{figure}

Fig.\ref{fig:nanofil}(a) schematically illustrates the Brillouin backscattering in a tapered optical nanofiber (ONF) which results from the coherent interaction between an optical pump wave depicted in blue and an acoustic wave (in purple). This optoacoustic interaction gives rise to an optical backscattered Stokes wave (in red) which is shifted in frequency from few GHz due to the Doppler effect. It is governed by both energy conservation as $\omega_a=\omega_p-\omega_s$, and momentum conservation with a phase-matching condition that reads as $\beta_a=\beta_p-\beta_s$, where $(\omega_i,\beta_i)_{i\in\{a,p,s\}}$ are the angular frequencies and propagation constants of the acoustic, the pump and Stokes modes, respectively. Unlike standard optical fibers, in sub-wavelength fibers, Brillouin backscattering has been shown to rely on several hybrid acoustic modes, including longitudinal, radial, torso-radial, and surface modes. \cite{beugnot_brillouin_2014,godet_brillouin_2017}. Among those acoustic modes, the torso-radial TR21 one is characterized by scalar potential and vector potential components with an orbital angular momentum with a topological charge $\ml = \pm{2}$, as shown in Figs.\ref{fig:nanofil}(b-e). However, since sub-wavelength optical fibers exhibit a slight ellipticity due to their manufacturing process based on the tapering of standard single-mode fibers \cite{lei_complete_2019}, the TR21 acoustic mode can be decomposed into two degenerated modes, which are denoted TR21$_+$ and TR21$_\times$ in Fig.\ref{fig:nanofil}(b) and (c). These two modes can also linearly be combined to two other vortex TR21$_+$ and TR21$_x$ modes, as in-phase and quadrature standing waves, with a topological charge $\ml = \pm{2}$.

To model the optoacoustic Brillouin interaction between the pump and these two acoustic modes, we used the elastodynamics equation driven by the electrostrictive stress, which can be written as\cite{beugnot_electrostriction_2012}
\begin{eqnarray}
	\rho\frac{\partial^2 u_i}{\partial t^2}-\left(c_{ijkl} u_{k,l}\right)_{,j}&=& -\left(\epsilon_0\chi_{ijkl}E^{(1)}_k E^{(2)*}_l\right)_{,j},
	\label{eq_1}
\end{eqnarray}
where $u_i$ are the displacement field components in three-dimensional space ($i\in\{x,y,z\}$), $c_{ijkl}$ are the elastic tensor components, $\rho$ is the silica material density and the index following a comma stands for the partial derivative with respect to that coordinate. On the right-hand side of Eq.(\ref{eq_1}), $E^{(1)}_k$ denotes the k-component of the optical pump electric field, $E^{(2)}_l$ being the $l$-component of the electric field of the optical Stokes signal, and $\chi_{ijkl}$ the electrostrictive tensor. We numerically solved Eq.(\ref{eq_1}) using a full-vectorial finite-element method (FEM) to get the displacement fields and the kinetic energy density. We specifically investigated different combinations of the pump and Stokes horizontal (H) and vertical (V) linear polarization. The results are summarized in Table \ref{tab:tab1}. We found that the displacement field that emerges is either the TR21$_+$ or the TR21$_\times$ mode, depending on the pump and Stokes linear polarization. 
\begin{table}[h]
    \centering
    \begin{tabular}{|c|c|c|}
    \hline
         Acoustic mode & Pump SOP & Stokes SOP \\
         \hline
         $\mathrm{TR21}_+$ & H&H \\
        $\mathrm{TR21}_+$ & V&-V \\ \hline
         $\mathrm{TR21}_\times$ & H&V \\        
        $\mathrm{TR21}_\times$ & V&H \\
        \hline
    \end{tabular}
    \caption{TR21 degenerated acoustic mode numerically computed Eq.(\ref{eq_1}) for different pump and Stokes states of polarization (SOP).}
    \label{tab:tab1}    
\end{table}
A straightforward result from Table \ref{tab:tab1} is that a linearly polarized pump light will be scattered into two orthogonally polarized Stokes beams by the TR21$_+$ and the TR21$_\times$ modes, respectively. In the case of spontaneous Brillouin scattering, the acoustic waves come from thermal agitation and thus the $\mathrm{TR21}_+$ and $\mathrm{TR21}_x$ waves are randomly phase-shifted. The state of polarization (SOP) of the Brillouin backscattered Stokes light, described by the Stokes vector $\vec{S_s}$ thus results from the incoherent superposition of both TR21$_+$ and TR21$_\times$ along the length L of the ONF. It can be written as follows:
\begin{eqnarray}
 \vec{S_s} &=& \int_{0}^{L}\left(M_{+}+M_{\Phi(z)} M_{\times}\right) \vec{S_p}\dd z, \label{eq:Pol1}
\end{eqnarray}
where $\vec{S_p}$ is the pump light Stokes vector and $M_{\Phi(z)}$ is a Mueller matrix built to apply a random phase shift $\Phi(z)$ distributed along the nanofiber z-axis. $M_+$ and $M_\times$ are the Mueller Matrices that describe how the polarization is affected when light is scattered by the TR21$_+$ or the TR21$_\times$ modes according to Table \ref{tab:tab1} \cite{parke_iii_optical_1949}. Note that $\vec{S_s}=(S_{s0},S_{s1},S_{s2},S_{s3})$ is the polarization Stokes vector, which is composed of four Stokes parameters.

In the following theoretical description and in the experimental section, we will adopt the convention of describing circular polarization from the point of view of the source (along the direction of the wave vector) \cite{ieeeconvention}. Two specific cases arise from Eq. \ref{eq:Pol1} depending on the pump SOP. First, if the pump light is linearly polarized, the Stokes SOP writes as
\begin{eqnarray}
 \vec{S_{s}}^{(1)} &=& \int_{0}^{L}\left(\vec{u}+M_{\Phi(z)} \vec{u_\bot}\right) \dd z,
 \label{eq:eq3}
\end{eqnarray}
where $\vec{u}$ is a linearly polarized SOP and $\vec{u_\bot}$ is the SOP orthogonal to $\vec{u}$. The superposition of two orthogonally polarized beams with random phase-shift leads to a scrambled polarization Stokes light, as observed in previous GAWBS experiments \cite{nishizawa_experimental_1995}. 

The second case deals with circular polarization that carries a spin angular momentum. When a circularly polarized pump wave $\vec{\nicefrac{R}{L}}$ is launched into the ONF (spin s = $\pm 1$), the two TR21$_{+/\times}$ contributions to the Brillouin scattering are such that the backscattered Stokes wave is fully polarized with an opposite circular polarization to the pump (s = $\mp 1$), as shown by the following expression,
\begin{eqnarray}
 \vec{S_{s}}^{(2)} 
 &=& \left(\int_{0}^{L}\left(1+M_{\pi/2+\Phi'(z)}\right) \dd z\right) \vec{\nicefrac{L}{R}},
\end{eqnarray}
where $\Phi'(z)$ is a random phase shift distributed along the nanofiber z-axis and $\vec{\nicefrac{L}{R}}$ stands for the left-handed or right-handed circular polarization Stokes vectors, respectively. In this specific case, the Stokes beam polarization is no longer scrambled. This has not been observed in GAWBS experiments but has recently been reported in chiral photonic crystal fibers \cite{zeng_stimulated_2022}. The easiest way to explain this circular polarization-handedness inversion is to consider the two TR+21 and TR-21 vortices modes with angular momentum $\pm 2$, as shown in Fig.\ref{fig:nanofil}(d) and (e). 
Using Table \ref{tab:tab1}, one can compute the outcome from the interaction between the TR+21 and TR-21 vortices and circular Pump SOPs. The results are shown in Table \ref{tab:tab2}. They can be interpreted as an optomechanical spin-orbit interaction in which the orbital angular momentum (OAM) $\ml$ of the acoustic vortices and spins of the pump and Stokes beams ($s_p$ and $s_s$) satisfy the conservation of total AM $\ml=s_p-s_s$, in addition to the phase matching condition $\beta_a=\beta_P-\beta_s$ and the energy conservation $\omega_a=\omega_p-\omega_s$. Since the two optical pump and Stokes waves are circularly polarized, they respectively carry SAM of $\pm1$ \cite{bliokh_polarization_2006}. On the acoustic side, a vortex with a topological charge $\ml$ is similar to an l-order Bessel beam owning both spin and orbital AM. Considering the cylindrical symmetry of the waveguide, the integration of the spin density over a volume makes the spin contribution to the total integral AM of the beam vanishing ($\langle S \rangle=0$). Consequently, the TR+21 and TR-21 acoustic waves exhibit AM with a purely orbital nature and the beams can be considered as acoustic vortices with azimuthal orders $\ml=\pm2$.

\begin{table}[ht]
    \centering
    \begin{tabular}{|c|c|c|c|}
    \hline
         Acoustic vortices & AOM ($\ml$)& Pump spin ($s_p$)& Stokes spin ($s_s$)\\
         \hline
         $\mathrm{TR}+21$ & +2& 1&-1 \\
        $\textcolor{red}{\mathrm{TR+21}}$ & \textcolor{red}{+2}& \textcolor{red}{-1}&\textcolor{red}{\sout{-3}} \\ \hline
         $\textcolor{red}{\mathrm{TR-21}}$ & \textcolor{red}{-2}& \textcolor{red}{1}&\textcolor{red}{\sout{3}}\\        
        $\mathrm{TR-21}$ & -2& -1&+1 \\
        \hline
    \end{tabular}
    \caption{TR21 degenerated acoustic modes excitation according to pump and Brillouin Stokes signal spin (numerical simulations). OAM, P, S correspond to Orbital Angular Momentum Pump and Stokes respectively. }
    \label{tab:tab2}    
\end{table}

Because the spin of light is restricted to $\pm 1$, the red lines in Table \ref{tab:tab2} show an additional outcome: A left circular polarized light can not be backscattered by a TR+21 acoustic mode and a right circular polarized light can not be backscattered by a TR-21 acoustic mode. This means that only the lines of Table \ref{tab:tab2} in black should occur, leading to a fully polarized Stokes beam even with a random phase relationship between the two vortex acoustic modes along the nanofiber.

\section{Brillouin scattering polarization-sensitive measurements}

\begin{figure}[ht]
    \centering
    \includegraphics[width=11 cm]{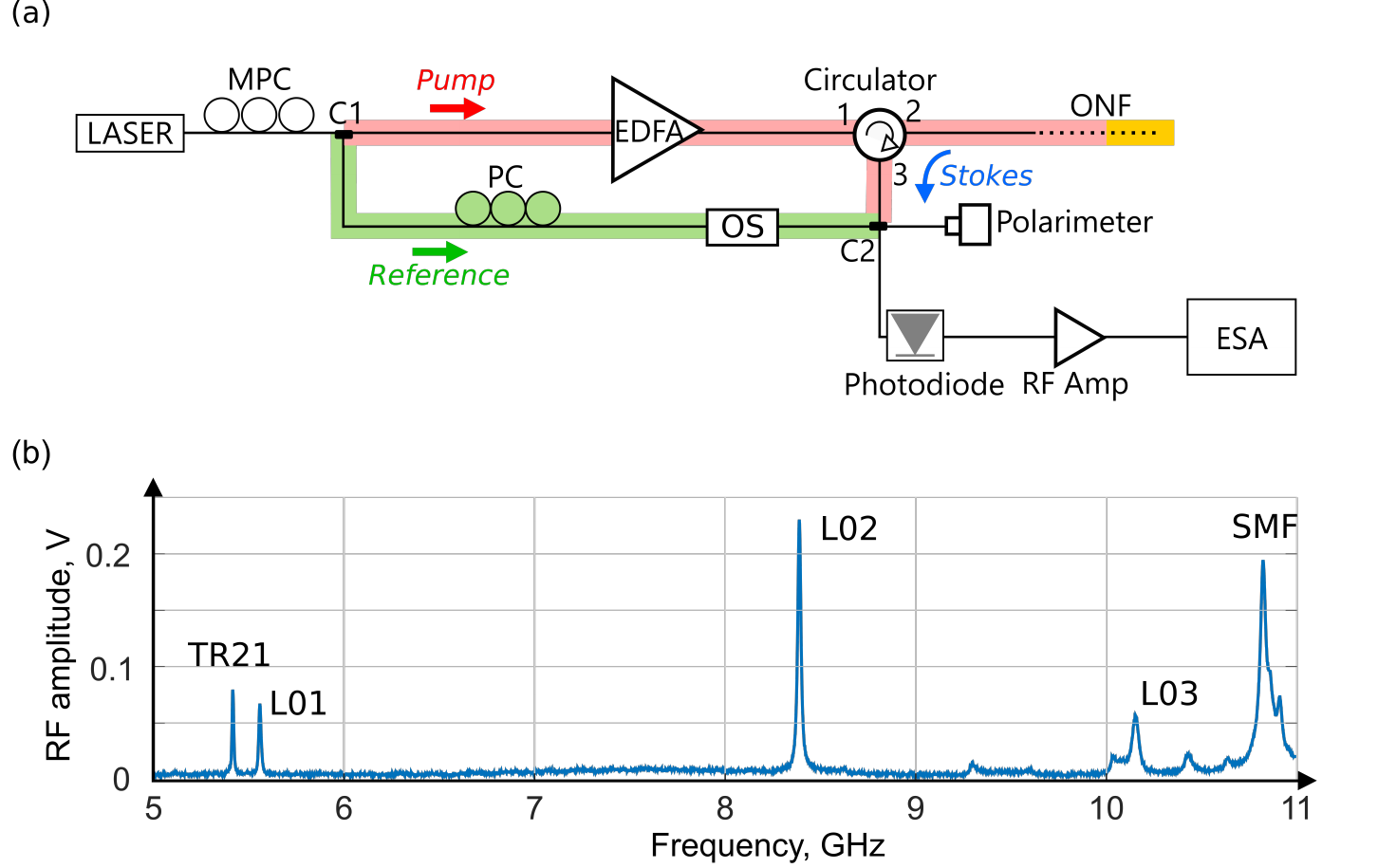}
    \caption{(a) Schematic illustration of the polarization-sensitive BS detection set-up. MPC: Motorized Polarization Controllers; EDFA: erbium-doped fiber amplifier; ONF: optical nanofiber; PC: manual polarization controllers; OS: optical switch; RF Amp: radio-frequency amplifier; ESA: electrical spectrum analyzer. (b) Measured ONF Brillouin spectrum with torso radial TR21 acoustic mode, L01, and L02 longitudinal acoustic mode signatures. The nanofiber has a 730\,nm diameter.} 
    \label{fig:setup}
\end{figure}

\begin{figure}[ht]
    \centering
    \includegraphics[width=0.7\textwidth]{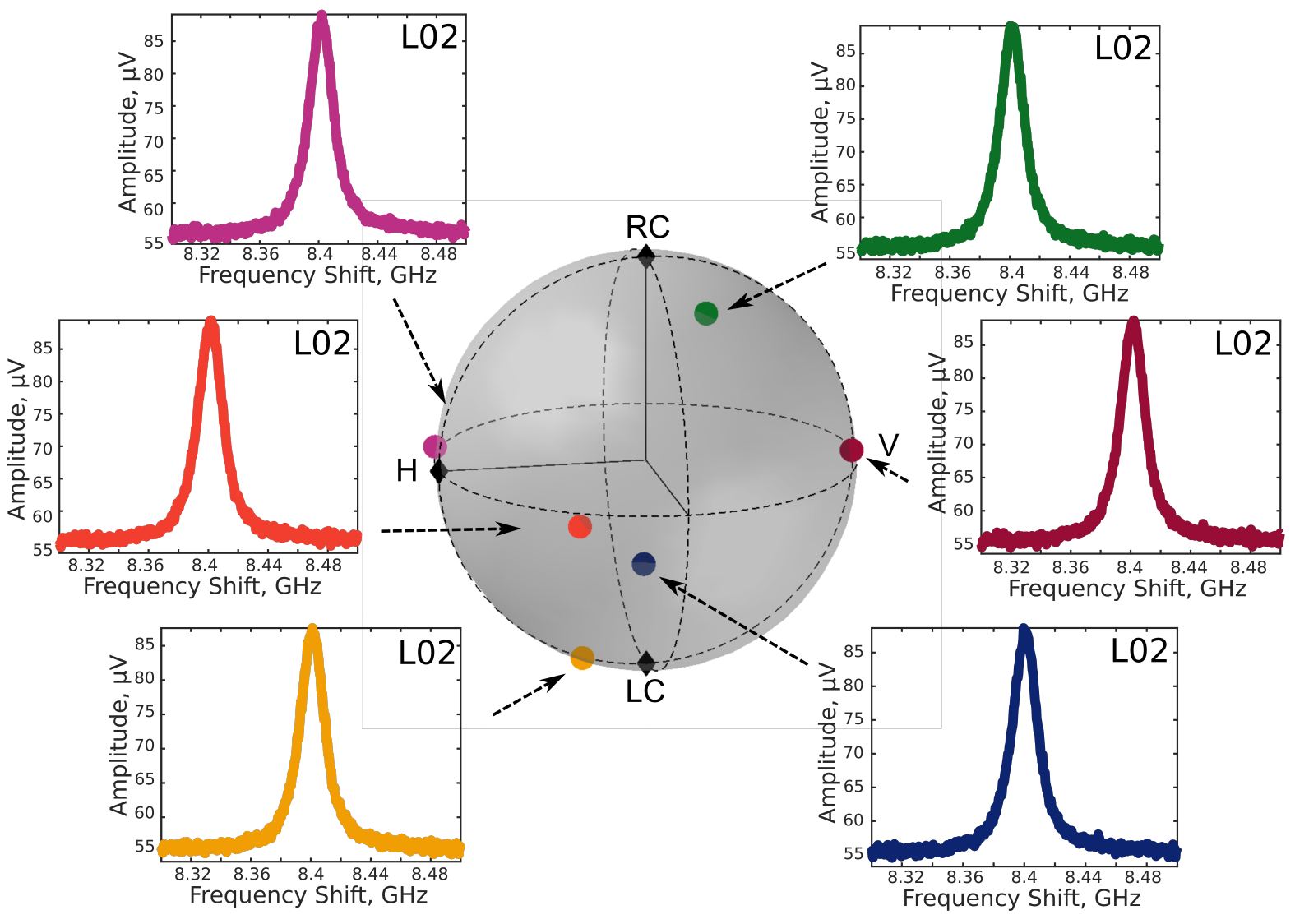}
    \caption{Experimental Brillouin spectra of the L02 acoustic mode resonance at 8.4 GHz from a 730~nm diameter silica nanofiber as a function of the input pump polarization indicated  on the Poincaré sphere.}
    \label{fig:L02measurements}
\end{figure}

The experimental setup used for polarization-sensitive measurements of Brillouin scattering and observation of the spin-orbit coupling-induced circular polarization reversal is depicted  in Fig.\ref{fig:setup}(a). We implemented a heterodyne detection technique from \cite{godet_brillouin_2017} for measuring the Brillouin backscattering combined with a polarization control and a polarimeter to observe the polarization dynamics. As a pump laser, we used a continuous-wave (CW) narrow-linewidth linearly-polarized DFB fiber laser at 1550 nm, which was sent to a motorized fiber polarization controller (MPC) to shape the input polarization state. It was then split using a directional fiber coupler (C1) into two beams. The top beam in pink serves as a pump for Brillouin backscattering, while the second beam in green is a reference beam or a local oscillator for heterodyne detection. The pump was then amplified using an erbium-doped fiber amplifier (EDFA) up to 33\,dBm and transmitted to the silica nanofiber through an optical circulator (port 2). The nanofiber was made from a standard single-mode silica fiber (SMF-28) using the tapering technique described in Ref.\cite{godet_brillouin_2017}. It has a diameter of 730 nm over a long uniform length of 10 cm. The Brillouin backscattered Stokes light from the ONF is then collected at port 3 (see blue arrow) and is mixed with the reference beam using a second fiber coupler (C2). The interference at the C2 output generates an optical beat signal that is further detected in the electrical domain using a fast photodiode. A polarimeter was placed on one of the two outputs of the coupler C2 to analyze the polarization of the beat signal. All fibers were maintained straight during the measurements to avoid any strain-induced polarization variation effects. The electrical signal was amplified and analyzed with an electrical spectrum analyzer (ESA), giving the Brillouin spectrum of the ONF, which is plotted in blue in Fig.\ref{fig:setup}(b). The two first peaks around 5.5\,GHz are the Brillouin signals from the torso-radial and longitudinal acoustic modes (TR21 and L01 modes), while the third peak at 8.4\,GHz comes from the hybrid acoustic L02 mode. The fourth main peak at 10.8\,GHz is due to the single-mode fiber (SMF) in the all-fiber setup, as already shown in Ref. \cite{godet_brillouin_2017}. Peak amplitude is greatest when both the reference beam (green) and the backscattered light (Stokes, blue) share the same SOP leading to strong interference. On the other hand, if both SOPs are orthogonal, the blue peaks vanish. The heterodyne measurement thus makes it possible to project the Stokes beam SOP onto the reference beam SOP by measuring the peak amplitudes. A manual polarization controller (PC) was added to the reference arm to control this projection. However, when the pump SOP is varied using the MPC, the reference beam SOP also varies due to the non-polarization maintaining fibers used in the setup. To minimize this effect, we adjust the polarization controller (PC) to maximize, for any given pump SOP, the interference between the reference beam and a beam that propagates along the pink path and is reflected at the cleaved output of the ONF (orange). This beam travels the same path as the Brillouin backscattered light except for the orange path. This setting allows us to compensate for any polarization rotation between the green and pink-orange paths. Since Brillouin scattering from an acoustic wave with no azimuthal order does not change the state of polarization \cite{shelby1985guided}, the amplitude of the hybrid acoustic wave (with both pressure and shear components) peak such as the L02 should be maximum whatever the SOP produced at the output of the MPC controller. This can be verified in Fig.\ref{fig:L02measurements} which shows various Brillouin spectra centered at the L02 mode frequency for six different input pump SOPs, pointed by a color marker on the Poincaré sphere. We notice in ref{fig:L02measurements} that the amplitude of the L02 peak is constant and independent of the pump SOP configuration. This indicates the effectiveness of our method for the alignment of the two optical paths.

\section{Experimental results}

\begin{figure}[ht!]
    \centering
    \includegraphics[width=.8\textwidth]{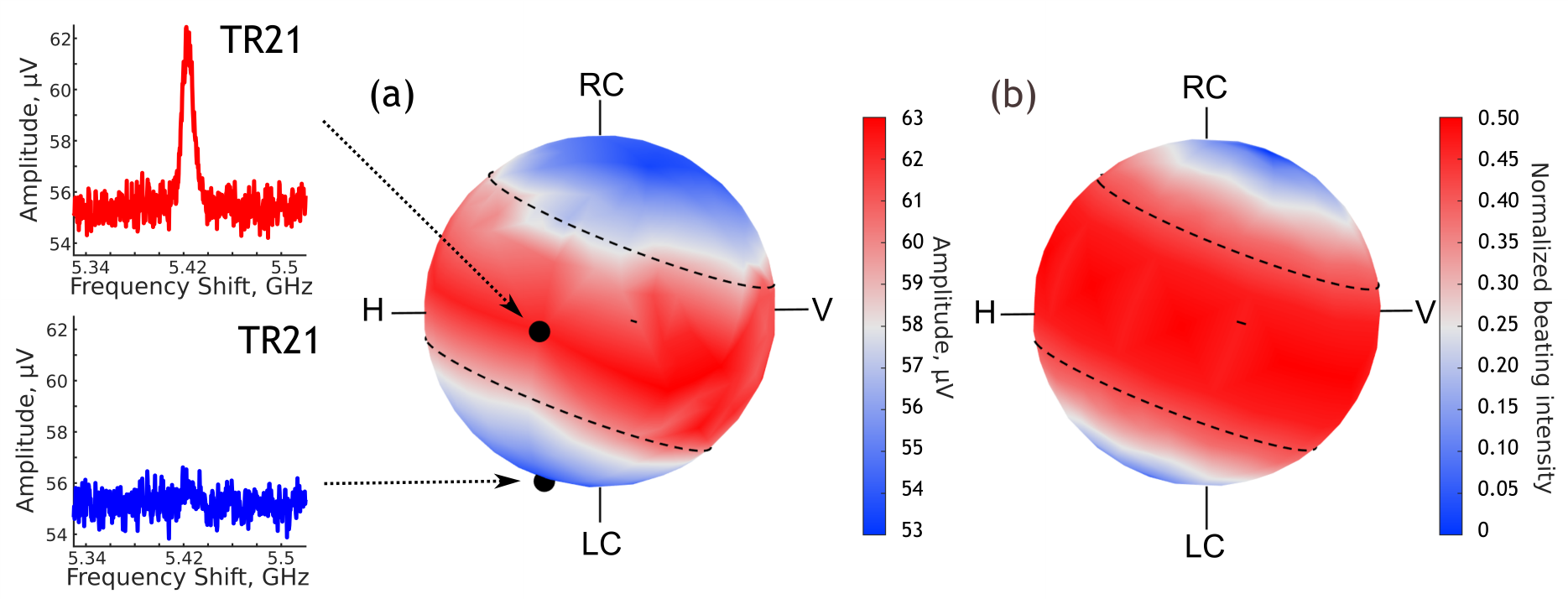}
    \caption{Polarization state dynamics of the TR21-mode induced Brillouin backscattering. Experimental (a) and numerical (b) TR21 peak intensity represented in false color on the Poincaré sphere describing the pump SOP. The two experimental spectra are extracted for the pump SOPs indicated with black markers.}
    \label{fig:spheres}
\end{figure}

With our setup, it is possible to go through several pump polarization states while projecting the SOP of the Stokes light onto the SOP of the pump. We, therefore, carry out the same experiment as before, this time reproducing the measured amplitude of the TR21 peak in false color, as shown in Fig. \ref{fig:spheres}(a) for each pump SOP on the Poincaré sphere. We can clearly observe an equatorial belt of higher intensity in red as opposed to two poles of minimum intensity (blue). The two markers on the sphere point to the spectra from the measurements for maximum (red, top) and minimum (blue, bottom) intensities.

\begin{figure}[ht!]
    \centering
    \includegraphics[width=.9\textwidth]{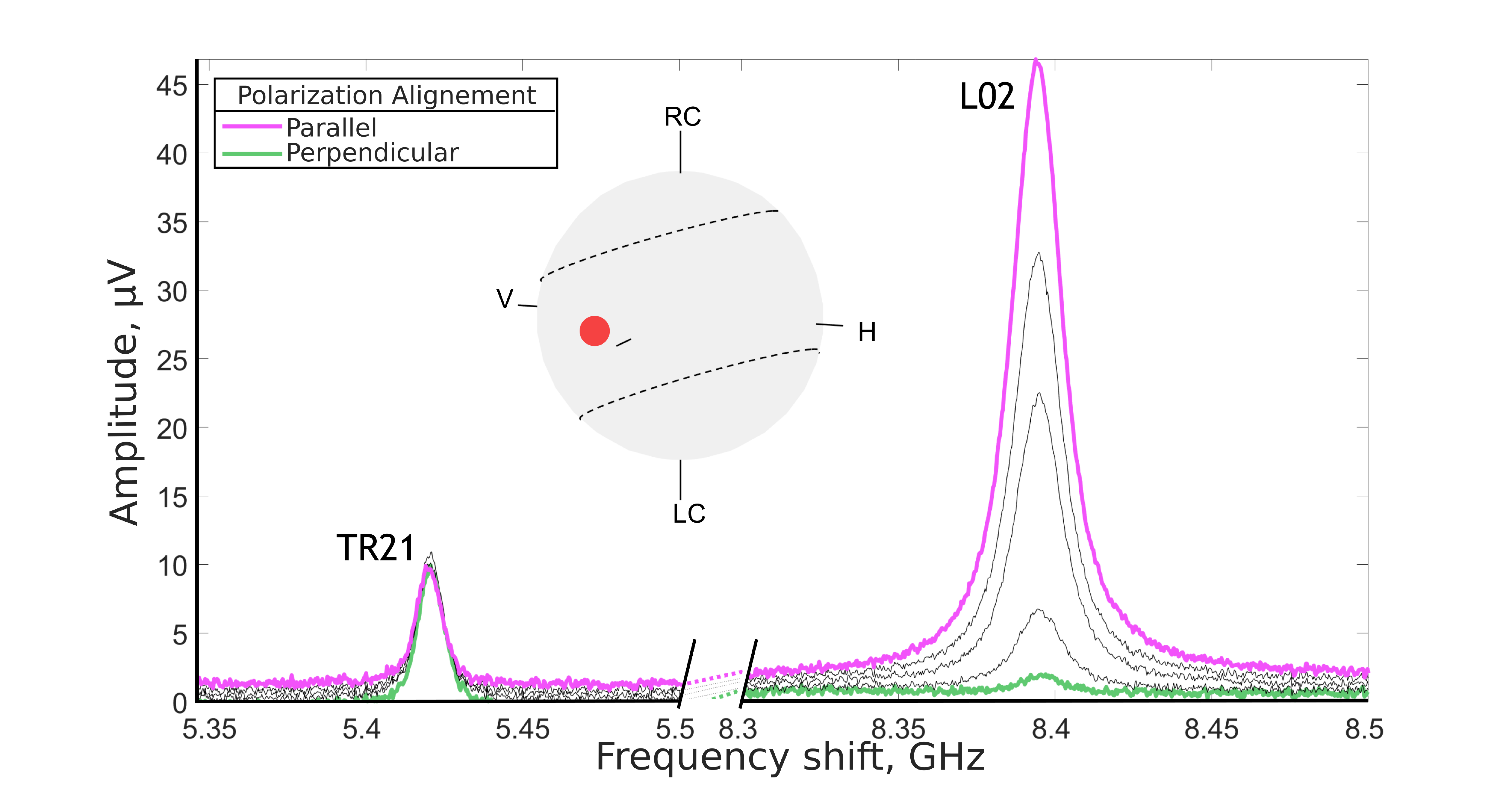}
    \caption{Experimental Brillouin spectra of the TR21 and L02 modes for different polarization of the reference beam. The linear pump SOP is depicted on the Poincaré sphere as a red dot and the reference beam polarization varies between parallel and perpendicular SOP. The TR21 signal appears depolarized.}
    \label{fig:unsensitivity}
\end{figure}

This confirms that, unlike the L02 mode originates from compression waves, the TR21 mode strongly affects the polarization of the Stokes wave. To interpret these results, we simulated the polarization dynamics on the sphere with the model introduced in the theoretical section. The numerical simulation is presented in Fig. \ref{fig:spheres}(b). As can be seen, we see as in the experimental figure a red equatorial belt delimited by black dashed lines. This corresponds to the area surrounding the linear polarization of the pump beam which gives rise to depolarized backscattered waves.Note that the blue color on \ref{fig:spheres}(b) corresponds to a zero beating intensity, as the reference and TR21 Stokes waves are cross-polarized, and the red color matches with the $0.5$ averaged intensity between the fully polarized reference wave and the depolarized TR21 Stokes wave.  Note also that the rotation of the poles with respect to the displayed reference frame (H,V,RC,LC) is due to the evolution of the Brillouin SOP between the center of the ONF and the polarimeter \cite{Joos:19}. 
To validate this interpretation, we fixed the pump polarization with a linear SOP, as depicted by the red marker in Fig.\ref{fig:unsensitivity} on the Poincaré sphere. The polarization of the reference beam was then rotated and we measured several Brillouin spectra, which are shown in grey in Fig. \ref{fig:unsensitivity}. On the right, the L02 peak near 8.4 GHz goes from a maximum intensity (pink) when the reference beam is co-polarized with the L02 Stokes wave to a near-zero minimum (green) when the reference wave is cross-polarized. We measured a high polarization extinction ratio (PER) of $\approx 38$\,dB, which confirms that the Brillouin Stokes wave backscattered by the L02 mode is well polarized. Conversely, we can clearly see in Fig. \ref{fig:unsensitivity} that there is almost no significant change of the Brillouin peak amplitude at 5.43 GHz due to the TR21 mode. The PER can be neglected ($\approx 0$\,dB). The TR21 peak is therefore insensitive to the polarization of the reference wave, which confirms the depolarization of the TR21 Stokes wave.

Since the linear polarization of the pump corresponds to the red belt in Fig. \ref{fig:spheres}, the blue poles are associated with the right-handed (RC) left-handed (LC) circular polarization, respectively. The blue areas on the Poincaré sphere thus reveal a TR21 peak close to zero, as shown by the blue spectrum on the left of Fig. \ref{fig:spheres}, due the absence of interference with the reference wave. This is only possible if the backscattered TR21 signal is fully polarized and its SOP is orthogonal to the pump SOP. Consequently, a left-handed circularly polarized pump wave will be backscattered into a right-handed circularly polarized Brillouin Stokes wave and vice versa. This is consistent with the theoretical results presented in Table \ref{tab:tab2} and this is due to the conservation of total angular momentum, which implies that the spin of the pump light couples with the orbital momentum of the acoustic vortex TR21 mode to generate a backscattered signal with a spin opposite to that of the pump wave.

\begin{figure}[t]
    \centering
    \includegraphics[width=.9\textwidth]{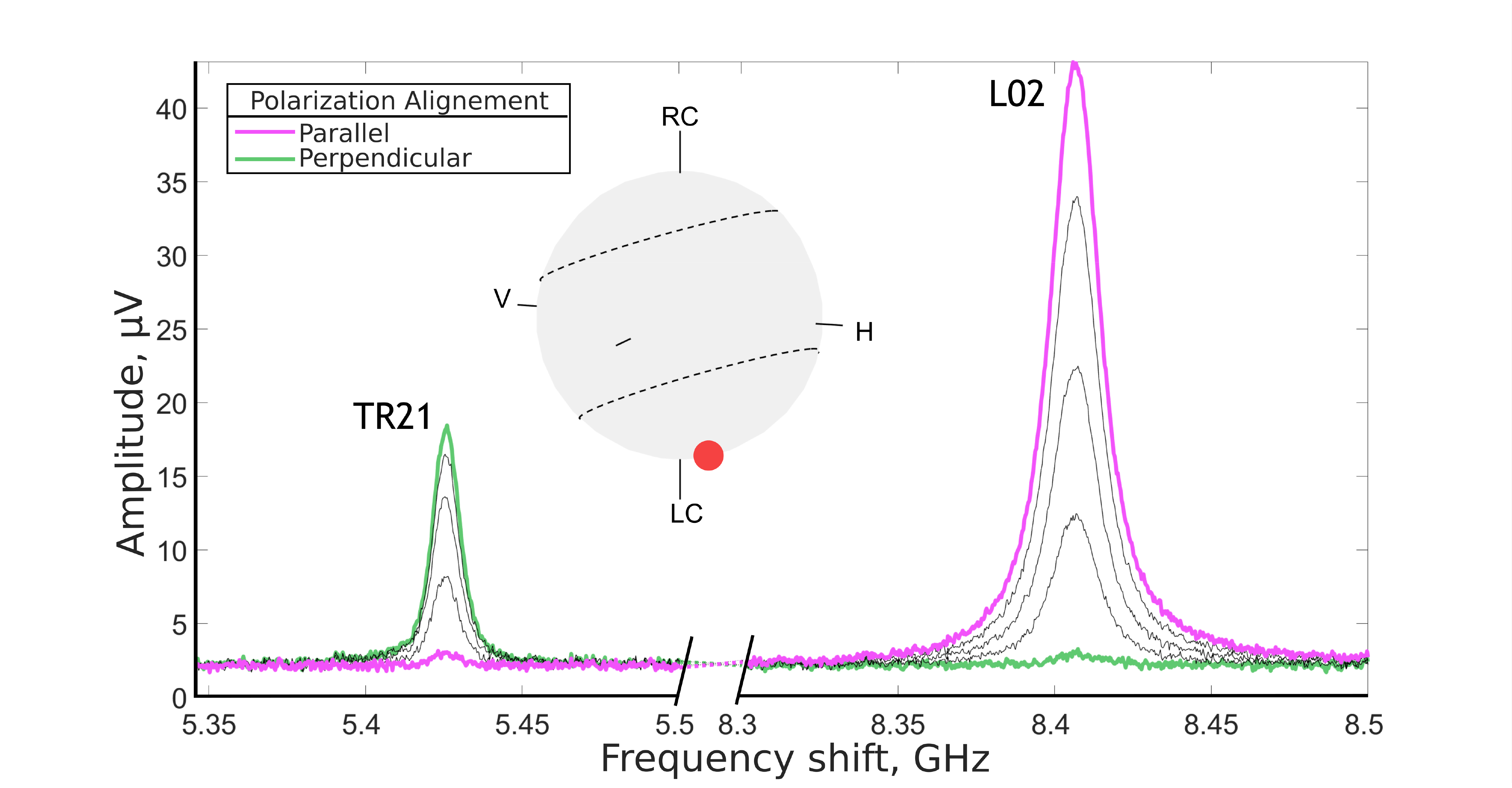}
    \caption{Polarization sensitive Brillouin TR21 backscattered  signal. The pump SOP is circular and depicted on the Poincaré sphere and the reference polarization varies between parallel and perpendicular SOP. The TR21 signal shows a polarization dependence.
    }
    \label{fig:sensitivity}
\end{figure}

To further validate our model, we then fixed the pump SOP on one of the blue poles (right circular polarization) of the Poincaré sphere, as shows the black dot in Fig. \ref{fig:spheres}(a). It is also depicted as a red dot in Fig. \ref{fig:sensitivity}. We performed the same measurements as before by varying the polarization of the reference beam and measuring the Brillouin spectra for the TR21 and the L02 modes, respectively. Results are shown in Figure \ref{fig:sensitivity}. The L02 peak shows similar high polarization-sensitive behavior as in Fig. \ref{fig:unsensitivity}, with an extinction ratio close to $32$\,dB. However, unlike the previous measurement, the TR21 now appears to be highly sensitive to the reference beam SOP. Indeed, a maximum peak amplitude is observed with the green curve while the peak is almost removed with the green curve, and the PER is as high as $26$\,dB. This confirms that the TR21 Stokes wave is well polarized and leads to strong interference. We can also clearly see in Fig. \ref{fig:sensitivity}  that, when the TR21 peak is maximal, the L02 peak is minimal and vice versa, meaning that the pump and Stokes waves are cross polarized. Since the pump wave is circularly polarized, this implies that, due to conservation of angular momentum,  a right/left-hand circularly polarized optical pump wave is backscattered into a left/right-hand circularly polarized Stokes light when interacting with an acoustic TR21 vortex wave with a topological charge of order $\pm 2$.

\section{Conclusion}

In this work, we theoretically and experimentally investigated the polarization properties of the backward Brillouin scattering in an optical nanofiber, with a special attention to the torso-radial vortex TR21 acoustic mode that has an angular orbital momentum. We predicted and observed two different dynamics depending on the polarization state of the pump beam. On one hand, when the pump beam is linearly polarized, the superposition of the incoherent contributions of the doubly degenerate TR21+ and TR21x modes induces a strong depolarization of the backscattered Stokes beam. On the other hand, when the pump beam is circularly polarized and thus carries a spin angular momentum ($\pm 1$), a spin-orbit interaction takes place and the angular momentum conservation implies that the backscattered Brillouin signal exhibits an opposite spin and thus a handedness circular polarization reversal. Those two regimes were modeled by a full-vectorial FEM of the optoacoustic interaction and then confirmed by experimental polarization-senstive measurements with very good agreement. The observation of this optomechanical spin-orbit coupling-induced circular polarization inversion is particularly reminiscent of Beth's historical experiment, while still finding great potential for future applications for all-optical information processing, all-optical memories, or in the development of non-reciprocal photonic components such a in-fiber isolators.

\section{Acknowledgements}
The authors acknowledge the financial support of EIPHI Graduate School (ANR-17-EURE-0002) and the Franche-Comté Region. They also thank Xinglin Zeng for helpful discussions.

\bibliographystyle{unsrt}
\bibliography{biblio1}

\end{document}